\documentclass[12pt]{article}
\begin{document}
\title{Qubit rotation in QHE}
\author{Dipti Banerjee$^*$\\
Department of Physics,Rishi Bankim Chandra College \\
Naihati,$24$-Parganas(N),Pin-$743 165$, West Bengal\\
INDIA}
\date{5.09.07}
\maketitle

\begin{abstract}
 In Quantum Hall effect the ground state wave function at $\nu=1$
 is the building block of all other states at different
 filling factors. It is developed by the entanglement of two spinors
 forming a singlet state. The inherent frustration visualized by
 the non-abelian matrix Berry phase is responsible for the quantum pumped charge
 to flow in the Hall surface. The Physics behind the Quantum Hall
states is studied here from the view point of topological quantum
computation.
\end{abstract}
Key words: spin echo, Berry phase.

 \vspace{3cm}PACS No:-$73.43$-f

\vspace{4cm}$*${This work of the author (Regular Associate) is
partially supported by ICTP,Trieste,Italy}.
 email: deepbancu@homail.com,dbanerje@ictp.it.

\section{ Introduction}
Entanglement is one of the basic aspects of quantum mechanics. It
was known long ago that quantum mechanics exhibits very peculiar
correlations between two physically distant parts of the total
system. Afterwards, the discovery of Bell's inequality (BI) [1]
showed that BI can be violated by quantum mechanics but has to be
satisfied by all local realistic theories. The violation of BI
demonstrates the presence of entanglement [2].
 The theorem of BI may be interpreted as incompatibility of requirement of locality
with the statistical predictions of quantum mechanics. So to study
the Bell state, the role of {\it local} spatial observations,
apart from spin correlations, should also be taken into account
[3]. This indicates
 that the spatial variation of a quantum
mechanical state would carry its memory through some geometric
phase known as Berry phase(BP) [4]. It is expected that the
influence of BP on an entangled state could be linked up with the
local observations of spins.

 To have a comprehensive view of the
quantum mechanical correlation between two spin $1/2$ particles in
an entangled state, we should take into account the role of the
Berry phase related to a spinor. This study belong to the field of
geometric quantum computation where noticeably the geometrical and
topological gates are resistant to local disturbances. In quantum
mechanical entanglement of two spin 1/2 particles the Berry phase
plays an important role during spin echo method [5]. Kitaev
described [6] the topological and quantum computer as a device in
which quantum numbers carried by quasiparticles residing in two
dimensional electron gas have long range Aharonov-Bohm (AB)
interactions between one another. These AB interactions are
responsible for nontrivial phase values during interwinding of
quasiparticles trajectories in course of time evolution of qubits
in Quantum Hall Effects (QHE). Quantization plays an important
role to realize the Physics behind different states of QHE from
the view point of Berry phase [7]. Recently we have studied the
rotation of a quantized spinor identified as qubit in presence of
magnetic field under the spin echo method [8]. We will aim at
understanding the rotation of QHE qubits specially in the lowest
Landau level $\nu=1$ and then parent states $\nu=1/m$ from the
view point of geometric quantum computation.

\section{Quantization of Fermi field and qubits of singlet states}

The quantization of Fermi field can be achieved assuming
anisotropy in the internal space through the introduction of
direction vector as an internal variable at each space-time point
[9]. The opposite orientations of the direction vector correspond
to particle and antiparticle. Incorporation of spinorial variables
$\theta(\bar{\theta})$ in the coordinate result the enlargement of
manifold from $S^2$ to $S^3$.This helps us to consider a
relativistic quantum particle as an extended one, where the
extension involves gauge degrees of freedom. As a result the
position and momentum variables  of a quantized particle becomes
\begin{equation}
Q_{\mu}=i\left( \frac{\partial}{{\partial p}_{\mu}} +
A_{\mu}\right),~~~~~~~~~P_{\mu}=i\left( \frac{\partial}{{\partial
q}_{\mu}} + \tilde{A_{\mu}}\right)
\end{equation}
where $q_{\mu}$ and $p_{\mu}$ are related to the position and
momentum coordinates in the sharp point limit and
$A_{\mu}(\tilde{A_{\mu}})$ are non-Abelian matrix valued gauge
fields belonging to the group $SL(2C)$.

In three space dimension, in an axis-symmetric system where the
anisotropy is introduced along a particular direction, the
components of the linear momentum satisfy a commutation relation
of the form [10]
\begin{equation}
[p_i, p_j] = i\mu \varepsilon_{ijk}\frac{x^k}{r^3}
\end{equation}
 Here $\mu$ corresponds to the measure of anisotropy and
behaves like the strength of a magnetic monopole. Indeed in this
anisotropic space the conserved angular momentum is given by
\begin{equation}
\vec{J}=\vec{r}\times\vec{p}- \mu\hat{r}
\end{equation}
with $\mu=0,\pm1/2,\pm1...$. This corresponds to the motion of a
charged particle in the field of a magnetic monopole. For the
specific case of $l=1/2,|m|=|\mu|=1/2$ for half orbital/spin
angular momentum, we can construct from the spherical harmonics
${Y_l}^{m, \mu}$, the instantaneous eigenstates $\left|\uparrow,t
\right\rangle$, representing the two component up-spinor as
\begin{eqnarray}
\left|\uparrow,t\right\rangle & = &{u\choose v} =
{{Y_{1/2}}^{1/2,1/2} \choose
{Y_{1/2}}^{-1/2,1/2}} \nonumber\\
&=& {\sin\frac{\theta}{2}\exp i(\phi-\chi)/2 \choose
\cos\frac{\theta}{2}\exp-i(\phi+\chi)/2}
\end{eqnarray}
and the conjugate state is a down-spinor given by
\begin{equation}
\left|\downarrow,t\right\rangle = {{-Y_{1/2}}^{-1/2,1/2} \choose
{Y_{1/2}}^{-1/2,-1/2}} = {-\cos\frac{\theta}{2}\exp i(\phi+\chi)/2
\choose \sin\frac{\theta}{2}\exp-i(\phi-\chi)/2}
\end{equation}
These two spinors (up/down) represent quantized fermi field
originated by an arbitrary superposition of elementary qubits
$\left|0\right\rangle and \left|1\right\rangle$ as for up spinor
\begin{equation}
\left|\uparrow,t\right\rangle
=\left(\sin\frac{\theta}{2}e^{i\phi}\left|0\right\rangle +
\cos\frac{\theta}{2}\left|1\right\rangle\right)e^{-i/2(\phi+\chi)}
\end{equation}
and the down spinor becomes
\begin{equation}
\left|\downarrow(t)\right\rangle =
(-\cos\frac{\theta}{2}\left|0\right\rangle +
\sin\frac{\theta}{2}e^{-i\phi}\left|1\right\rangle)e^{i/2(\phi+\chi)}
\end{equation}


The states $\left|\uparrow,t\right\rangle$ and
$\left|\downarrow,t\right\rangle$ can be generated by the unitary
transformation matrix $U(\theta,\phi,\chi)$ [11]
\begin{equation}
U(\theta,\phi,\chi)=
\pmatrix{{\sin\frac{\theta}{2}e^{i/2(\phi-\chi)}~~~~~~~~~
-\cos\frac{\theta}{2}e^{i/2(\phi+\chi)}} \cr
\cos\frac{\theta}{2}e^{-i/2(\phi+\chi)}~~~~~~~~~
{\sin\frac{\theta}{2}e^{-i/2(\phi-\chi)}}}
\end{equation}
in association with the basic qubits $\left|0\right\rangle$ and
$\left|1\right\rangle$
\begin{eqnarray}
\left|\uparrow,t\right\rangle =
U(\theta,\phi,\chi)\left|0\right\rangle ,
\left|\downarrow,t\right\rangle=U(\theta,\phi,\chi)\left|1\right\rangle
\end{eqnarray}

Over a closed path, the single quantized up spinor acquires the
geometrical phase [8]
\begin{eqnarray}
\gamma_{\uparrow}&=& i\oint\left\langle\uparrow,t\right|\nabla\left|\uparrow,t\right\rangle.d\lambda\\
&=& i\oint\left\langle0\right|U\dag dU\left|0\right\rangle.d\lambda\\
 &=& i\oint A_{\uparrow}(\lambda)d\lambda\\
 &=& \oint L^\uparrow_{eff} dt\\
 &=& \frac{1}{2}(\oint d\chi-\cos\theta\oint d\phi)\\
 &=& \pi(1 - \cos\theta)
\end{eqnarray}
representing a solid angle subtended about the quantization axis.
For the conjugate state the Berry phase over the closed path
becomes
\begin{equation}
\gamma_{\downarrow} = -\pi(1-\cos\theta)
\end{equation}
 The fermionic or the antifermionic
nature of the two spinors (up/down) can be identified by the
maximum value of topological phase $\gamma_{\uparrow/
\downarrow}=\pm \pi$ at an angle $\theta=\pi/2$. For $\theta=0$ we
get the minimum value of $\gamma_{\uparrow}=0$ and at $\theta=\pi$
no extra effect of phase is realized.

It can be verified that this Berry phase remains the same if we
neglect the overall phase $e^{\pm i(\phi-\chi)/2}$ from the
quantized spinors as in eqs.(6) and (7) respectively. The
identical value of Berry phase $\gamma_{\uparrow/\downarrow}$ in
both the approaches is only possible if we consider no local
frustration in the spin system otherwise the conflict between the
parameters of quantized spinor will cause to have different BP
[12].

In the language of quantum computation, the rotation of qubit or
quantized spinor can be studied well in the background of
geometric phase. Any electronic state at any instant can be
written as a linear combination of the instantaneous eigenstates.
\begin{equation}
\left|\Psi(t)\right\rangle = c_1 (t)\left|\Phi_1 (t)\right\rangle
+ c_2 (t)\left|\Phi_2 (t)\right\rangle
\end{equation}
In a cyclic change of the time period T, the instantaneous basis
states $\left|\Phi_1 (T)\right\rangle$ and $\left|\Phi_2
(T)\right\rangle$ might return to their initial states
$\left|\Phi_1 (0)\right\rangle$ and $\left|\Phi_2
(0)\right\rangle$ where the coefficients $c_1 (T)$ and $c_2 (T)$
may not. This doubly degenerate energy level, a $2\times2$ matrix
Berry phase $\Phi_c$ connects the final amplitudes- $c_1 (T), c_2
(T)$ with the initial amplitudes $c_1 (0), c_2 (0)$
\begin{equation}
{c_1 (T) \choose c_2 (T)} = \Phi_c {c_1 (0) \choose  c_2 (0)}
\end{equation}
 Hwang et.al [13] pointed out that this non abelian
 matrix Berry phase is responsible for pumped charges where the
 charge transport in a cycle of the pump in a jth optimal channel
 becomes
 \begin{equation}
Q_j = -i/2\pi \oint \left\langle\psi_j\middle|d\psi_j\right\rangle
\end{equation}
This charge will be only of topological in nature during transport
of qubits, if the influence of dynamical phase can be eliminated.
Spin echo method is a popular technic for this removal of
dynamical phase where two cyclic evolutions are applied on a
spinor with the second application followed by a pair of fast
$\pi$ transformations. Vedral et.al.[14] showed the application of
spin echo to a spinor (eq.(6)).
\begin{eqnarray}
\left|\uparrow\right\rangle&\longrightarrow^{C_R}&
 e^{i(\delta_\uparrow-\gamma)}\left|\uparrow\right\rangle
{\longrightarrow}^\pi e^{i(\delta_\uparrow - \gamma)}
\left|\downarrow\right\rangle  \nonumber\\
 &\longrightarrow^{C_L}& e^{i(\delta_\uparrow +\delta_\downarrow -
2\gamma)}\left|\downarrow\right\rangle {\longrightarrow}^\pi
e^{i(\delta_\uparrow +\delta_\downarrow -
2\gamma)}\left|\uparrow\right\rangle
\end{eqnarray}
Here ${\longrightarrow}^{C_R}$ introduces the dynamical and
geometrical phases, $\delta_\uparrow$ and $\gamma_\uparrow$
through right cyclic evolution of $\left|\uparrow\right\rangle$
spinor respectively. Similar phases of opposite orientations are
developed by $\longrightarrow^{C_L}$. Referring back to eqs.(15)
and (16), we see that $\gamma_\uparrow$=$\gamma$ and
$\gamma_\downarrow$=$-\gamma$ for $\gamma=\pi(1-cos\theta)$. Thus
two cyclic evolutions accompanied by two $\pi$ rotations eliminate
the net dynamical phases doubling the geometric phase of the
original state (up/down spinor) according to.
\begin{equation}
\left|\uparrow\right\rangle\longrightarrow e^{2i
\gamma_\uparrow}\left|\uparrow\right\rangle,
  \left|\downarrow\right\rangle\longrightarrow e^{2i \gamma_{\downarrow}}\left|\downarrow\right\rangle
\end{equation}
For two half periods of spin echo rotation we have
\begin{equation}
\left|\uparrow\right\rangle\longrightarrow
 e^{i \gamma_\uparrow}\left|\uparrow\right\rangle,
 \left|\downarrow\right\rangle\longrightarrow e^{i\gamma_{\downarrow}}\left|\downarrow\right\rangle
\end{equation}
where the total effect of dynamical phase disappear. The spin echo
method is very fruitful [15] in the construction of two qubit
through rotation of one qubit (spin 1/2) in the vicinity of
another.
 Incorporating the spin-echo for half period (as in eqn.22)
 we find the antisymmetric Bell's state after one cycle $(t=\tau)$,
\begin{equation}
\left|\Psi_- (t=\tau)\right\rangle =
\frac{1}{\sqrt{2}}(e^{i\gamma_\uparrow}\left|\uparrow\right\rangle_1\otimes
\left|\downarrow\right\rangle_2 -
e^{-i\gamma_\uparrow}\left|\downarrow\right\rangle_1
\otimes\left|\uparrow\right\rangle_2)
\end{equation}
and symmetric state becomes
\begin{equation}
\left|\Psi_+ (t=\tau)\right\rangle =
\frac{1}{\sqrt{2}}(e^{-i\gamma_\uparrow}\left|\uparrow\right\rangle_1\otimes
\left|\downarrow\right\rangle_2 +
e^{i\gamma_\uparrow}\left|\downarrow\right\rangle_1
\otimes\left|\uparrow\right\rangle_2)
\end{equation}
 where $\gamma_\downarrow = -\gamma_\uparrow =- \gamma$.
Splitting up these above two eqs.(23) and (24) into the symmetric
and antisymmetric states and rearranging we have
\begin{eqnarray}
\left|\Psi_+\right\rangle_\tau &=& \cos\gamma \left|\Psi_+\right\rangle_0 - i \sin\gamma \left|\Psi_-\right\rangle_0 \\
\left|\Psi_-\right\rangle_\tau &=& i \sin\gamma
\left|\Psi_+\right\rangle_0 + \cos\gamma
\left|\Psi_-\right\rangle_0
\end{eqnarray}
the doublet acquiring the matrix Berry phase-$\Sigma$ as rotated
from  $t=0$ to $t=\tau$.
\begin{equation}
{\left|\Psi_+\right\rangle \choose \left|\Psi_-\right\rangle}_\tau
= \Sigma {\left|\Psi_+\right\rangle \choose
\left|\Psi_-\right\rangle}_0
\end{equation}

\begin{equation}
\Sigma = \pmatrix{{\cos\gamma~~~-i\sin\gamma}\cr
{i\sin\gamma~~~~\cos\gamma}} = \cos 2\gamma
\end{equation}
This non-abelian matrix Berry phase $\Sigma$ is developed from the
abelian Berry phase $\gamma$. For $\gamma=0$ there is symmetric
rotation of states, but for $\gamma=\pi$ the return is
antisymmetric as the values of $\Sigma$=I and -I (where I=identity
matrix) respectively.

 The instantaneous quantum
state can be represented by the linear combination of degenerate
symmetric and antisymmetric states. Symmetric state will return to
antisymmetric state over one half period of spin echo apart from a
matrix valued Berry phase [16]. It may be noted that two half
period rotations will complete one spin echo resulting the return
of the state to itself apart from a geometrical phase factor.
\begin{equation}
{\left|\Psi_+\right\rangle \choose
\left|\Psi_-\right\rangle}_{2\tau}
=\pmatrix{{\cos\gamma~~~-i\sin\gamma}\cr
{i\sin\gamma~~~~\cos\gamma}}\pmatrix{{\cos\gamma~~~i\sin\gamma}\cr
{-i\sin\gamma~~~~\cos\gamma}}\nonumber {\left|\Psi_+\right\rangle
\choose \left|\Psi_-\right\rangle}_0
\end{equation}
Following the notion of one complete spin echo here, the state
$\left|\Psi_+\right\rangle_{T=2\tau}$ also return to its initial
state ${\left|\Psi_+\right\rangle}_0$ apart from the phase
$\cos2\gamma$.
\begin{equation}
{\left|\Psi_+\right\rangle \choose
\left|\Psi_-\right\rangle}_{2\tau}=\pmatrix{{\cos 2\gamma~~~0}\cr
{0~~~~\cos2\gamma}}{\left|\Psi_+\right\rangle \choose
\left|\Psi_-\right\rangle}_0
\end{equation}
In any even number of half period $\tau$, the symmetric state will
return to itself apart from Berry phase factor with increased
power of $\cos 2\gamma$.
\begin{equation}
{\left|\Psi_+\right\rangle\choose\left|\Psi_-\right\rangle}_{2n\tau}=\pmatrix{{\cos
2\gamma~~~0}\cr{0~~~~\cos 2\gamma}}^n{\left|\Psi_+\right\rangle
\choose \left|\Psi_-\right\rangle}_0
\end{equation}
where $n=1,2,3...$ are natural integers. For odd number of the
half periods rotations there will be mixture of both the states.
 On the other hand with the value of $\gamma=\pi$,the
symmetric/antisymmetric state remains same after one rotation.

In this connection we have shown recently [8] that the singlet
state between two spinors at a particular instant is connected
with the singlet state of elementary qubits $\left|0\right\rangle$
and $\left|1\right\rangle$ and the Berry phase of the initial
antisymmetric Bell's state is $\gamma_{ent}=\pi(1+\cos2\theta)$
where if we introduce spin echo in the two qubit then the
topological phase $\Sigma = cos 2\gamma$ is of matrix valued.

 By varying the magnetic field angle
$\theta:0\longrightarrow\pi/3\longrightarrow\pi/2$, the Berry
phase(BP) of a qubit changes to,
$\gamma:0\longrightarrow\pi/2\longrightarrow\pi$, that in turn
change the two qubit BP, $\Sigma: I\longrightarrow
\sigma^y\longrightarrow -I$. This explains the physics behind the
change from the antisymmetric Bell singlet state $\Psi_-$ to the
symmetric Bell state $\Psi_+$ and back to $\Psi_-$. We will now
proceed to apply the above idea of entanglement in the field of
Quantum Hall effect to study the state formation from one filling
factor to another in the light of Geometric Quantum Computation.

 \section{\bf Qubit formation of Quantum Hall
state }
 Quantum Hall effect shows a prominent appearance of
quantization of Hall particles involving gauge theoretic extension
of coordinate by $C_{\mu}\epsilon SL(2C)$ visualized by the field
strength $F_{\mu\nu}$ acting as background external magnetic
field. It is noted that the gauge field theoretic extension for a
Fermi field associated with the direction vector $\xi_\mu$
attached to the space-time point $x_\mu$ results the field
function $\phi(x_\mu, \xi_\mu)$  describing a particle moving in
an anisotropic space [7].

The external magnetic field introduces frustration in the Hall
system. We have considered a two-dimensional frustrated electron
gas of N particles on the spherical surface of a three dimensional
sphere of large radius R in a strong radial (monopole) magnetic
field. In such a 3D anisotropic space we can construct the
N-particle wave-function from the spherical harmonics
${Y_l}^{m,\mu}$ with $l=1/2$, $|m|=|\mu|=1/2$ (when the angular
momentum in the anisotropic space is given by eq.(3)). With the
description of a two component up spinor
$\left|\uparrow\right\rangle = {u \choose v}$ as in eq.(6) we can
construct the $N$ particles wave function of Hall states
\begin{equation}
{\Psi_{N_\uparrow}}^{(m)}= \prod(u_i v_j - u_j v_i)^m
\end{equation}
for parent states $m=1/\nu$ where $\nu$ is the Landau filling
factor and  this $m=J_{ij}= J_i+J_j$ is the two particle angular
momentum equivalent to $m= \mu_i + \mu_j=2\mu$ (when $i=j$).
 Similar manner the same Hall state with opposite
 polarization can be constructed by using the
down spinor $\left|\downarrow\right\rangle={\tilde{v} \choose
\tilde{u}}$
\begin{equation}
{\Psi_{N_\downarrow}}^{(m)}=
\prod(\tilde{u}_i\tilde{v}_j-\tilde{u}_j\tilde{v}_i)^m
\end{equation}
Here the two states ${\Psi_{N_{\uparrow}}}^{(m)}$ and
${\Psi_{N_{\downarrow}}}^{(m)}$ belong to the same parent filling
factor but with opposite polarization of the spinors.

The above states are grouped into a family depending on the value
of $m$. With $m=3$ the states are the same family of the Laughlin
$\nu=1/3$ state etc. In the light of Jain [17] that regarding the
filling factor the IQHE of composite fermions are the FQHE of
fermions, any FQHE state can be expressed in terms of the IQHE
state. It seems that for LLL $\nu=1$, IQHE state $\Phi_1(z)$
\begin{equation}
\Phi_1 (z)=(u_i v_j - u_j v_i)
\end{equation}
is the basic building block for constructing any other IQHE/FQHE
state. The lowest level Hall state $\Phi_1(z)$ has a similarity
with two-qubit singlet state formed by a pair of one qubit states.

There is a deep analogy between FQHE and superfluidity [18]. The
ground state of anti-ferromagnetic Heisenberg model on a lattice
introduce frustration giving rise to the resonating valence
bond(RVB) states corresponding spin singlets where two
nearest-neighbor bonds are allowed to resonate among themselves.
It is suggested that RVB states [6] is a basis of fault tolerant
topological quantum computation. Since these spin singlet states
forming a RVB gas is equivalent to fractional quantum Hall fluid,
its description through quantum computation will be of ample
interest.

This resonating valence bond(RVB) where two nearest-neighbour
bonds are allowed to resonate among themselves has equivalence
with entangled state of two one-qubit. The antisymmetric Hall
state $\Phi_1 (z)$  for $\nu=1$ is formed as one spinor at ith
site rotating with Berry phase $\gamma=\pm i\pi$ in the vicinity
of another at jth site captures the image of spin echo
\begin{eqnarray}
\left|\Phi_1 (z)\right\rangle
&=&\frac{1}{\sqrt{2}}(\left|\uparrow\right\rangle_1~~\left|\downarrow\right\rangle_1)\pmatrix
{{0~~~-e^{-i\pi}}\cr{e^{i\pi}~~~0}}{\left|\uparrow\right\rangle_2
\choose\left|\downarrow\right\rangle_2}\\
 &=&(\left|\uparrow\right\rangle_1~~~\left|\downarrow\right\rangle_1)\pmatrix
{{0~~~1}\cr{-1~~~0}}{\left|\uparrow\right\rangle_2 \choose \left|\downarrow\right\rangle_2}\\
&=&\frac{1}{\sqrt{2}}(\left|\uparrow\right\rangle_1~~\left|\downarrow\right\rangle_2
-\left|\downarrow\right\rangle_1~~\left|\uparrow\right\rangle_2)\\
&=&\left|\Psi_-\right\rangle
\end{eqnarray}
Due to symmetry, the singlet state can be written on any basis
with the same form. We can rotate the spin vector by an arbitrary
angle $\theta$ with the following transformation.
\begin{equation}
{\left|\uparrow\right\rangle \choose
\left|\downarrow\right\rangle}=\pmatrix{{\sin\theta e^{i\phi}
~~~\cos\theta}\cr{-\cos\theta~~~~\sin\theta
e^{-i\phi}}}{\left|0\right\rangle \choose \left|1\right\rangle}
\end{equation}
 The Quantum Hall systems are so highly
frustrated that the ground state $\Phi_1(z)$ is an extremely
entangled state visualized by the formation of antisymmetric
singlet state between a pair of $i,j$th spinors in the Landau
filling factor $(\nu=1)$.
\begin{eqnarray}
\Phi_1 (z) &=& \pmatrix{{u_i~~~~~u_j}\cr{v_i~~~~~v_j}}=(u_i v_j -
u_j v_i) \nonumber\\
&=& (u_i~~~~v_i)\pmatrix {{0~~~~~1}\cr{-1~~~~~0}}{u_j \choose v_j}
\end{eqnarray}
 We identify this two qubit singlet state as Hall qubit constructed
 from the up-spinor shown in the previous section
 \begin{equation}
\Phi_1 (z) = \left\langle\uparrow_i\right|\pmatrix
{{0~~~~~1}\cr{-1~~~~~0}}\left|\uparrow_j\right\rangle =
\left\langle0\right|U_i\dag \pmatrix
{{0~~~~~1}\cr{-1~~~~~0}}U_j\left|0\right\rangle
\end{equation}
The down spinor can construct the opposite polarization of Hall
qubit
\begin{equation}
\Phi_1 (\tilde{z})=(\tilde{u}_i\tilde{v}_j-\tilde{u}_j\tilde{v}_i)
\end{equation}
that has a similar representation as eq.(41)
\begin{equation}
\Phi_1 (\tilde{z}) = \left\langle1\right|U\dag \pmatrix
{{0~~~~~1}\cr{-1~~~~~0}}U\left|1\right\rangle
\end{equation}
Now these two Hall qubits of two opposite polarizations
representing the state of same lowest Landau level $\nu=m=1$ will
automatically generate two respective non-abelian Berry
connections. The Hall connection for up spinor becomes
\begin{equation}
B_{\uparrow}=\Phi_1 (z)^* d \Phi_1 (z)
 = \pmatrix{{{u_i}^*~~~~{v_i}^*}\cr{{u_j}^*~~~~{v_j}^*}}
\pmatrix{{du_i~~~~du_j}\cr{dv_i~~~~dv_j}}
\end{equation}
and similarly for down-spinor
\begin{equation}
\tilde{B_{\downarrow}}=\Phi_1 (\tilde{z})^* d \Phi_1 (\tilde{z})
 =\pmatrix{{\tilde{u_i}^*~~~~\tilde{v_i}^*}\cr{\tilde{u_j}^*~~~~\tilde{v_j}^*}}
\pmatrix
{{d\tilde{u_i}~~~~d\tilde{u_j}}\cr{d\tilde{v_i}~~~~d\tilde{v_j}}}
\end{equation}
The non-abelian nature of the connection or Berry phase on the
Hall surface for the lowest Landau level LLL ($\nu=1$) will remain
if $i\neq j$.
\begin{eqnarray}
B_{\uparrow}&=& \pmatrix{{(u_i^*du_i +v_i^*dv_i)~~~~(u_i^*du_j
+v_i^*dv_j)}\cr{(u_j^*du_i +v_j^*dv_i)~~~~(u_i^*du_i +v_i^*dv_i)}}\nonumber\\
&=&\pmatrix{{\mu_i~~~\mu_{ij}}\cr{\mu_{ji}~~~~\mu_j}}
\end{eqnarray}
This is visualizing the spin conflict during parallel transport
leading to matrix Berry phase. In the light of Hwang et.al [13]
our realization includes that in Quantum Hall effect this
non-abelian matrix Berry phase is responsible for the charge flow
by pumping. In this QHE matrix Berry phase
\begin{equation}
{\gamma^H}_{\uparrow}=\pmatrix{{\gamma_i~~~~\gamma_{ij}}\cr{\gamma_{ji}~~~~\gamma_j}}
\end{equation}
 $\gamma_i$ and $\gamma_j$ are the BPs for the ith
and jth spinor as seen in eq. (15) and the off-diagonal BP
$\gamma_{ij}$ arises due to local frustration in the spin system.
 Over a closed period $t=\tau$ the QHE state $\Phi_1(z)$ at $\nu=1$
filling factor will acquire the matrix Berry phase.
\begin{equation}
\left\langle\Phi_1 (z)\right|_\tau =
e^{i{\gamma^H}_{\uparrow}}\left\langle\Phi_1 (z)\right|_0
\end{equation}
Berry connection gets modified as the quantum state differ after
one rotation. Usually when any state changes by
 $$\left|\psi'\right\rangle =\left|\psi\right\rangle e^{i\Omega(c)}$$
 the corresponding changed gauge
 becomes $$ {A_{\psi}}'=A_{\psi}+id \Omega(c)$$
 provided $\left\langle\psi\middle|\psi\right\rangle=1$.
 We have pointed out earlier [19] that each Quantum Hall state for
a particular filling factor has its distinct Berry phase. Hence BP
is constant for a filling factor. The rotation shifts the BP from
ground to excited level once. With these ideas we have the
topological phase difference between the first excited and the
ground state  acquired by the rate of change of Berry phase
\begin{equation}
\Gamma^1 -\Gamma^0 = i\oint {\left\langle\Phi_1
(z)\right|d\gamma^H/d\lambda \left|\Phi_1 (z)\right\rangle}_0
d\lambda
\end{equation}
 The rotation of singlet state by 'n' number of turns will be
\begin{equation}
{\left\langle\Phi_1 (z)\right|^n}_{\tau} =
e^{i{n\gamma^H}_{\uparrow}}{\left\langle\Phi_1 (z)\right|^n}_0
\end{equation}
 where $n=1,2,3..$ are the natural numbers associated with the
number of rotations of the singlet states. We should point out
here that the antisymmetric nature of FQHE states would be
visualized through the rotation of singlet states. This
automatically imposes the following constraint in the topological
phase
\begin{equation}
e^{{in\gamma}^H}=e^{im\pi}=-1,
\end{equation}
$$ for~~~~ \left\langle\Phi_1 (z)\right|_{n\tau} =-\left\langle\Phi_1(z)\right|_0$$
where $m=1,3,5..$ being the odd numbers to maintain the
antisymmetric nature of wave function. So any number of rotations
of the matrix Berry phase lead to odd multiple of $\pi$ angles
provided the every state remains antisymmetric. It seems that BP
act as a local order parameter of QHE states.
\begin{equation}
{\left\langle\Phi_1 (z)\right|}^{m\pi/\gamma}_{\tau} =
e^{im\pi}{\left\langle\Phi_1 (z)\right|}^{m\pi/\gamma}_0
\end{equation}
Earlier we showed [7] that the Berry phase for $\nu=1/m$ state is
$\gamma== m\pi\theta=2\pi\mu\theta$ where $\theta$ is a coupling
constant.This motivated us to write
\begin{equation}
{\left\langle\Phi_1
(z)\right|}_{\tau}=e^{im\pi\theta}{\left\langle\Phi_1
(z)\right|}_0
\end{equation}
This makes the experimental observation of parent state in FQHE at
$m=odd(3,5,7)$ more transparent. It also shows that the
topological phase is responsible for controlling the statistics of
the Hall state. In absence of frustration, the role of
 matrix Berry phase is trivial. In other words
 $\gamma_{ij}$ becomes zero leading to diagonal matrix Berry phase
 provided the two particle have identical $\theta$ and $\phi$
values.
\begin{equation}
{\gamma^H}_{\uparrow}=\pi(1-\cos\theta_i)\pmatrix{{1~~~~0}\cr{0~~~~1}}
\end{equation}
The non-abelian matrix Berry phase in Quantum Hall effect is
originated due to the frustration offered by the magnetic field
and the disorder of spins. In the absence of local frustration
(latter) this complexity of connection will be removed. We would
like to mention that spin echo between two single qubit has the
equivalence of RVB state in FQHE and topological quantum
computation with BP is responsible for the formation of higher
states considering the Hall qubit at $\nu=1$ as a building block
of any QHE state.

\section{Discussion}

In this paper we have studied the Physics behind the singlet state
entangled by the two qubits where one is rotating in the field of
the other with the Berry phase only. This image of spin echo has
been reflected in the field of Quantum Hall effect. The Hall state
for the lowest Landau level at $\nu=1$ is highly frustrated. They
are the singlet states identified as the Hall qubit, the building
block of other higher IQHE/FQHE states at different filling
factors. These states have matrix Berry phase which are
responsible for pumped charge flow. In other words the Berry phase
acts as a local order parameter of singlet states. Further we
pointed out that the antisymmetric nature of $\nu=1/m$ FQHE states
depend on their acquired Berry phase. Since these spin singlet
states forming a RVB gas is equivalent to fractional quantum Hall
fluid, the description of background Physics through quantum
computation will be of ample interest. We will proceed to study
the hierarchies of FQHE in the light of quantum communication in
the future.
\\ \\ \\

{\bf Acknowledgements:}\\
This work is partly supported by The Abdus Salam International
Center for Theoretical Physics, Trieste, Italy. The author would
like to acknowledge the help from all the works cited in the
reference. Moreover the author is thankful to the referees and
Editor of my paper in PRB for fruitful comments.

\pagebreak


\end{document}